\documentclass[prb,twoside,twocolumn,showpacs]{revtex4}
\usepackage{graphicx}
\usepackage{amsmath}
\usepackage{amssymb}

\begin{document}

\title{The different development of the anisotropic upper critical field in MgB$_2$ by aluminum and carbon doping}
\author{M. Angst}
  \email[Email: ]{angst@ameslab.gov}
\author{S.~L. Bud'ko}
\author{R.~H.~T. Wilke}
\author{P.~C. Canfield}
\affiliation{Ames Laboratory and Department of Physics and
Astronomy, Iowa State University,Ames,IA 50011,USA}

\date{\today}
\begin{abstract}
The temperature dependence of the upper critical field, $H_{c2}$,
for both field directions in partially substituted polycrystalline
MgB$_2$ was determined. Whereas the suppression of $T_c$ is
similar for aluminum and carbon substituted samples, $H_{c2}$ is
affected by the substitution in profoundly different ways. In the
case of Al substitution changes can tentatively be described by
intrinsic effects (shift of the Fermi level). In the C substituted
samples, $H_{c2}$ is increased drastically, and extrinsic effects
(disorder) have to play a major role. The strong contrast between
the two substitutions is discussed, taking into account three
relevant scattering rates.
\end{abstract}
\pacs{74.25.Op, 74.20.De, 74.25.Ha, 74.70.Ad}
\maketitle

%\section{Introduction}
\label{intro} An unusual temperature dependence of the anisotropic
upper critical field, $H_{c2}$, is one of the major consequences
of two-band superconductivity as realized in magnesium diboride
MgB$_2$.\cite{Angst03Nova} To further explore properties of a
given compound it is often helpful to consider the effects of
partial chemical substitutions. In the case of MgB$_2$, partial
substitutions with many elements have been attempted, but only two
elements are widely recognized to enter the structure: aluminum
replacing magnesium,\cite{Slusky01} and carbon replacing
boron.\cite{Ribeiro03}
%Recently a
%substitution of scandium for magnesium has also been
%proposed.\cite{Agrestini04}

Both substitutions dope the material with additional electrons,
which should similarly affect the superconducting properties, at
least to the extent that a rigid band approximation works.
According to the detailed band structure calculations
\cite{Kortus01,An01,Mazin03} electron doping most drastically
affects the $\sigma$ bands, which are nearly filled. Furthermore,
any partial substitution by small amounts of an additional element
increases chemical disorder, leading to increased scattering. In
the two-band superconductor MgB$_2$, at least three different
scattering rates have to be distinguished,\cite{Mazin02} and the
different substitution sites Mg (by Al) and B (by C) are likely to
influence these scattering rates in drastically different ways. In
general, the upper critical field, $H_{c2}$, is influenced by
electron-phonon coupling (EPC), Fermi velocities, and by the mean
free path, $\ell$. EPC and Fermi velocities are intrinsic
properties altered by electron doping, while $\ell$ is a function
of scattering. It will be interesting to compare the doping and
temperature dependence of $H_{c2}$ with substitutions on either
the Mg or the B site. This may help in separating electron doping
and scattering effects of partial substitutions.

In the case of the B site substitution by carbon a number of
studies have presented measurements of $H_{c2}$ on polycrystalline
materials \cite{Ribeiro03,Holanova04,Wilke04} and in a limited
range on single
crystals.\cite{Lee03,Masui04,noteOhmichi04,Pissas04,Puzniak04,Kazakov04}
All studies agree in significant enhancements of $H_{c2}$, and the
studies on single crystals also indicate a decrease of the
$H_{c2}$ anisotropy $\gamma_H \equiv H_{c2}^{\|ab} /
H_{c2}^{\|c}$. Fewer $H_{c2}$ studies exist for Mg site
substitution by aluminum, and there is a considerable spread of
given values between them.\cite{Putti04b,Putti04,Kang04b}

Here, we present a comparison of $H_{c2}^{\|c}$ and
$H_{c2}^{\|ab}$ measured with the same technique on aluminum and
carbon substituted polycrystalline MgB$_2$ with various
substitution levels, prepared from the same Mg and (partly) B
starting materials with similar procedures. As a function of
electron doping, a similar decrease of both the transition
temperature, $T_c$, and the upper critical field anisotropy,
$\gamma_H$, contrasts with the dramatically different development
of the magnitude of $H_{c2}$: Whereas the behavior of the upper
critical field in the case of Al substitution can be understood as
resulting from a shift of the Fermi level, an increase in
scattering has to be taken into account to explain the large
increase of $H_{c2}$ upon C substitution. We briefly discuss this
different effect on scattering by C and Al substitution.

%\section{Experimental}

We investigated carbon substituted polycrystalline MgB$_2$ samples
prepared in two different ways: We synthesized
Mg(B$_{0.9}$C$_{0.1}$)$_2$ at $1200^{\circ}{\rm C}$ using Mg and
B$_4$C as starting materials, as described in Ref.\
\onlinecite{Ribeiro03}. For low C substitution levels, filaments
already studied in Ref.\ \onlinecite{Wilke04} were ground to
powder. In the case of aluminum substitution, chemical
inhomogeneities are difficult to avoid. Inhomogeneities lead to
transition broadening detrimental particularly to the
determination of $H_{c2}^{\|c}$. In order to maximize sample
homogeneity, we tried several techniques, including pre-alloying
Mg and Al, and using AlB$_2$ and AlB$_{12}$ as Al source. However,
the best results were obtained with a two-step synthesis at high
temperatures. First, synthesis at constant temperatures from
$1000$ to $1200^{\circ}{\rm C}$ from the elements in
stoichiometric quantities for up to $10$ days, followed by cooling
to room temperature in streaming water produced material with
large inhomogeneities as visible in x-ray diffraction pattern and
particularly in the superconducting transition. In order to
improve homogeneity, we finely ground and thoroughly mixed the
products of the above synthesis, pressed them into pellets, and
then heated them for a second time to $1200^{\circ}{\rm C}$ for
$10$ days.

%\section{Results and Discussion}
\label{res}

\begin{figure}[tb]
\includegraphics[width=0.95\linewidth]{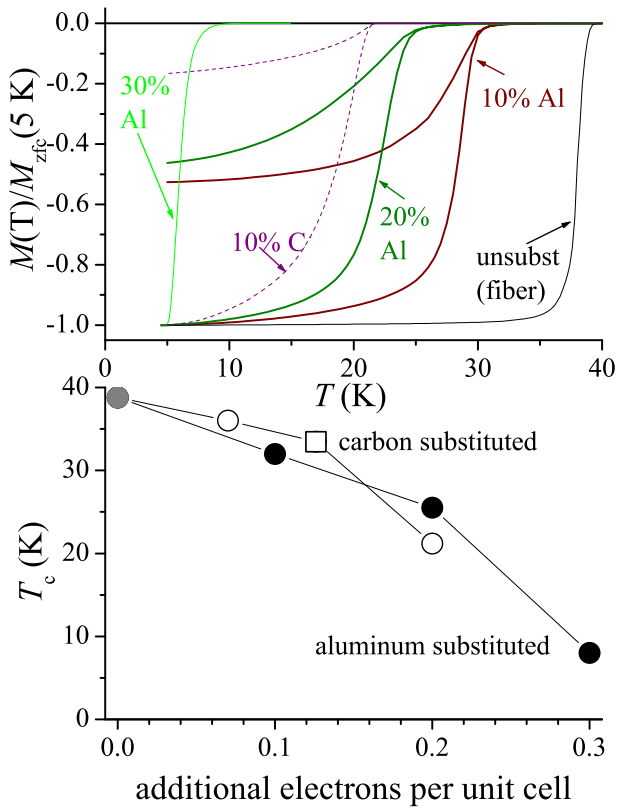}
\caption{a) Magnetization $M$ vs temperature $T$ in $20\,{\rm Oe}$
(both zero field cooled and field cooled) of Al or C substituted
polycrystalline MgB$_2$ samples b) Transition temperature $T_c$ vs
number of additional electrons per unit cell due to Al (full
symbols) or C (open symbols) substitution. The open square is from
Ref.\ \onlinecite{Puzniak04}.} \label{Fig1}
\end{figure}

Powder x-ray measurements on MgB$_2$ samples substituted with up
to 20\% Al indicate no phase separation, shifts in the lattice
parameters close to literature values,\cite{Putti03} and a
moderate peak broadening suggesting small variations in the Al
content throughout the samples. The broadening becomes significant
for Mg$_{0.7}$Al$_{0.3}$B$_2$. For this composition, additional
small peaks suggest the presence of MgB$_2$ and MgAlB$_4$ minority
phases. The superconducting transition in zero field was measured
resistively (on the $60$ to $70$\% dense pellets) and by
magnetization measurements in $20\,{\rm Oe}$ (after powdering the
samples and mixing with epoxy). The magnetization measurements on
isolated powder particles makes any $T_c$ variations within the
sample well visible as an onset broadening [Fig.\ \ref{Fig1}a)].
The effective $T_c$ was defined as the crossing point of the
steepest slope of the field cooled $M(T)$ with the $M=0$ axis
[Fig.\ \ref{Fig1}b)]. The height of the ``sliver'' in the onset
above this temperature is a measure for the amount of material
with higher $T_c$ due to locally less Al substitution. Such a
variation is present in Al substituted samples, but rather small.

\begin{figure}[tb]
\includegraphics[width=0.95\linewidth]{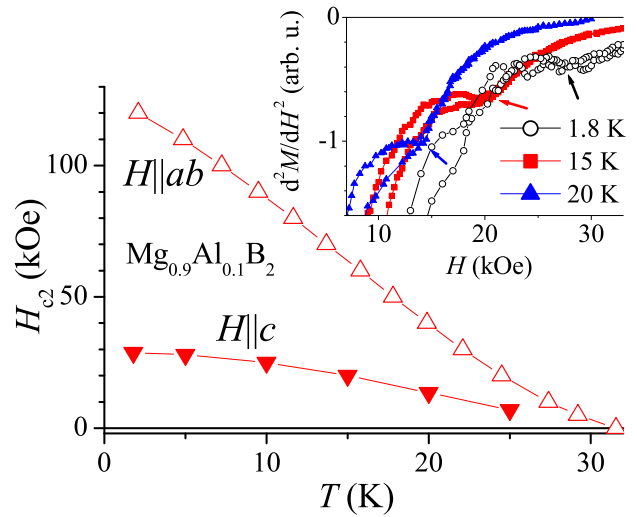}
\caption{Upper critical field $H_{c2}$ of
Mg$_{0.9}$Al$_{0.1}$B$_2$. Inset: Second derivative ${\rm d}^2 M /
{\rm d} H^2$ of the magnetization as a function of field $H$ at
different temperatures on Mg$_{0.9}$Al$_{0.1}$B$_2$. The location
of the minimum upper critical field (cf.\ Ref.\
\onlinecite{Budko02}), i.e., $H_{c2}^{\|c}$, is indicated by
arrows.} \label{Fig2}
\end{figure}

The polycrystalline upper critical field, which corresponds to
$H_{c2}^{\|ab}$, was determined from resistivity and magnetization
in applied fields up to $140\,{\rm kOe}$ and $70\,{\rm kOe}$,
respectively. In the overlapping field region, the results agree
within error bars. The results on the C substituted samples are
also in agreement with the results of Refs.\
\onlinecite{Holanova04} and \onlinecite{Wilke04}. The ``minimum
upper critical fields'' ($H_{c2}^{\|c}$) were established with a
method developed by Bud'ko and coworkers.\cite{Budko01b,Budko02}
On unsubstituted polycrystalline MgB$_2$, this method yielded
similar results\cite{Budko02} on the temperature dependent
anisotropy as measurements\cite{MgB2anisPRL02} performed on single
crystals. Materials were ground to a fine powder and mixed with
epoxy. The minimum upper critical field is then visible as
pronounced features in the derivatives of the magnetization as a
function of temperature or field (see inset of Fig.\ \ref{Fig2}).
As an example, the resulting upper critical field of
Mg$_{0.9}$Al$_{0.1}$B$_2$ is shown in Fig.\ \ref{Fig2}. For this
sample, $H_{c2}^{\|c}(0) \simeq 29\,{\rm kOe}$ is slightly higher
than $H_{c2}^{\|c}(0) \simeq 25\,{\rm kOe}$ measured with the same
method on an unsubstituted sample.\cite{Budko02} The significance
of the increase is questionable, taking into account that even on
unsubstituted single crystals from the same source
$H_{c2}^{\|c}(0)$ with values of $28\,{\rm kOe}$ (Ref.\
\onlinecite{MgB2anislow}) up to $31\,{\rm kOe}$ (Ref.\
\onlinecite{MgB2anisPRL02}) were observed. We note that our result
is significantly lower than the value of $52\,{\rm kOe}$ reported
for a Mg$_{0.88}$Al$_{0.12}$B$_2$ single crystal.\cite{Kang04b}
The upper critical field parallel to the layers on the other hand,
$H_{c2}^{\|ab}(0) \simeq 127{\rm kOe}$, is significantly lower
than corresponding measurements on unsubstituted MgB$_2$. Our
result is moderately higher than the results reported in Refs.\
\onlinecite{Putti04}and \onlinecite{Kang04b}, significantly higher
than those reported in Ref.\ \onlinecite{Putti04b}. Parts of the
discrepancies might be related to different amounts of impurity
scattering (see discussion below), whereas other parts may
originate from different inhomogeneities in the Al distribution.

\begin{figure}[tb]
\includegraphics[width=0.95\linewidth]{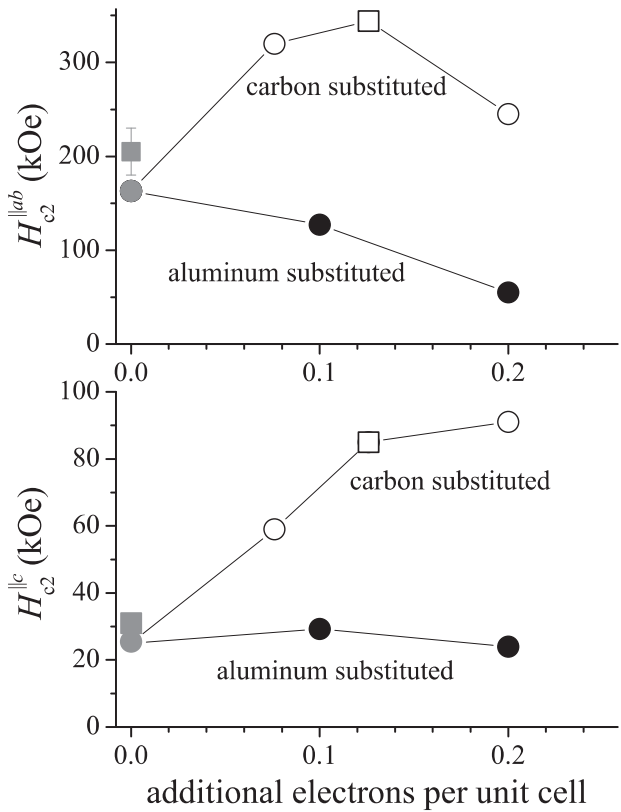}
\caption{Zero temperature upper critical field vs number of
additional electrons per unit cell due to Al (full symbols) or C
(open symbols) substitution. Grey symbols are unsubstituted
MgB$_2$. a) $H_{c2}^{\|c}(0)$. b) $H_{c2}^{\|ab}(0)$. Squares in
both panels are single crystal results from Refs.\
\onlinecite{Puzniak04} and \onlinecite{MgB2anisPRL02}, open
circles in panel a) are from Refs.\ \onlinecite{Holanova04} and
\onlinecite{Wilke04}.} \label{Fig3}
\end{figure}

The analysis on the aluminum doped samples with up to 20\% Al
substitution\cite{noteLi02} yields the following picture: Al
substitution first slightly increases, then slightly decreases
$H_{c2}^{\|c}$ (essentially constant), whereas $H_{c2}$ parallel
to the layers monotonically decreases (Fig.\ \ref{Fig3}, closed
symbols). The decrease of $H_{c2}^{\|ab}$ is roughly linearly,
extrapolating to 0 for $\sim $30\% Al substitution.

The decrease of $H_{c2}^{\|ab}$ and the almost constant behavior
of $H_{c2}^{\|c}$ suggests that disorder may not be important in
determining the $H_{c2}$ development with Al substitution. In
unsubstituted MgB$_2$, clean limit (two-band) theoretical
calculations\cite{Miranovic03,Dahm03} compare rather favorably
with experimental data.\cite{MgB2PhyC,Dahm03} These calculations
as well as phenomenological considerations\cite{Angst03Nova}
indicate that in the low temperature limit the $\pi$ bands are not
important for determining $H_{c2}$: $H_{c2}(0)$ is mostly
determined by the $\sigma$ bands, as first suggested by Bud'ko and
coworkers.\cite{Budko01b} The upper critical field is related to
the coherence length, $\xi$, through $H_{c2} \propto \xi^{-2}$. In
the clean limit at zero temperature, ignoring the difference
between GL and BCS coherence length, $\xi$ is related to the
superconducting gap and Fermi velocities by $\xi \propto v_F/
\Delta$, and the Fermi velocity anisotropy determines the
anisotropy of $\xi$. Here, $v_F$ ($\Delta$) is defined as the
root-mean-squared wave vector dependent Fermi velocity
(superconducting gap), averaged over the Fermi surface (in the
case here over the $\sigma$ sheets of the Fermi surface). We may
then approximate
\begin{equation}
H_{c2}^{\|c}(0) \propto  \left ( \Delta_{\sigma}(0)  /
v_{F,\sigma}^{\|ab} \right ) ^2,\: \gamma_H (0) =
v_{F,\sigma}^{\|ab} / v_{F,\sigma}^{\|c}. \label{xigap}
\end{equation}

Apart from disorder effects due to the partial substitution, Al
doping modifies charge distribution and decreases the lattice
constants ($c$ in particular).\cite{Pena02} The {\em main} effect,
however, is to dope the system with additional electrons,
resulting in a shift of the Fermi level, $E_F$, to higher
energies. For substitution levels well below 30\%, where $E_F$
reaches the $\sigma$ band edge at the $\Gamma$ point,\cite{Pena02}
the changes in the electronic structure are well approximated
within a rigid band model. The increase of $E_F$ decreases the
density of states (DOS) at $E_F$ and modifies the band averaged
Fermi velocities, primarily in the $\sigma$ bands.\cite{Mazin03}
For moderate substitution levels, the out-of-plane $\sigma$ Fermi
velocity, $v_{F,\sigma}^{\|c}$, remains approximately constant,
whereas the in-plane $\sigma$ Fermi velocity,
$v_{F,\sigma}^{\|ab}$, substantially decreases. According to Eq.\
(\ref{xigap}) this lowers the $H_{c2}$ anisotropy and increases
$H_{c2}^{\|c}$. However, the decreased DOS at $E_F$ weakens the
electron-phonon coupling, resulting in decreased superconducting
gaps and $T_c$ (c.f.\ Fig.\ \ref{Fig1}). The additional effects of
the substitution on the phonons\cite{Renker02} complicate the
theoretical analysis of the development of $\Delta_{\sigma}$ and
$\Delta_{\pi}$ with substitution level, and experimental reports
on the gap development are sparse as of
yet.\cite{Putti03,Putti04,Gonnelli04b} As an approximation, we can
use the experimental values of the transition temperature shown in
figure \ref{Fig1}, and assume $\Delta_{\sigma} \propto T_c$.

The combination of the decreased $\Delta_{\sigma}$ and also
decreased $v_{F,\sigma}^{\|ab}$ results in little change of
$H_{c2}^{\|c}$ as estimated by Eq.\ (\ref{xigap}), in accordance
with the experiment. The calculation also yields the substantial
decrease of $H_{c2}^{\|ab}$ and of the anisotropy $\gamma_H$
observed experimentally. For the out-of-plane upper critical
fields of Mg$_{1-x}$Al$_{x}$B$_2$, a similar, but slightly more
detailed analysis was recently presented by Putti and
coworkers.\cite{Putti04} The fact that the experimental
development of the upper critical field can be accounted for by
the clean limit formula (\ref{xigap}) clearly suggests that
effects of increased scattering are not relevant in our samples of
Mg$_{1-x}$Al$_{x}$B$_2$ at low substitution levels. This is
different from recently presented results on single crystal
samples, where scattering in the $\pi$ bands seems to be
larger.\cite{Kang04b}

\begin{figure}[tb]
\includegraphics[width=0.95\linewidth]{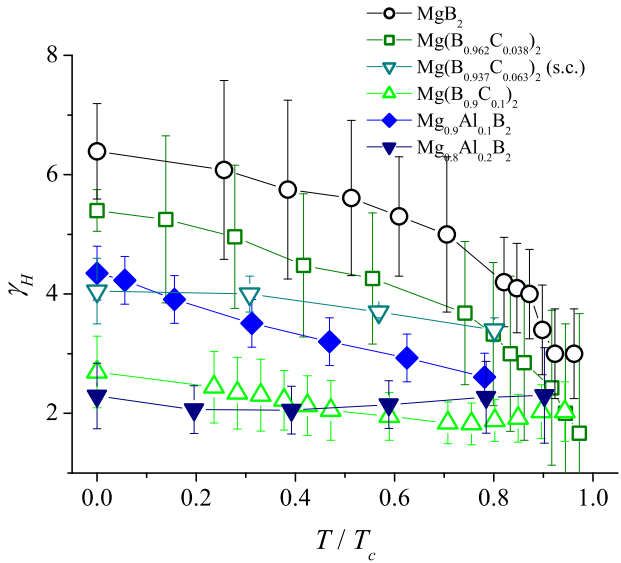}
\caption{Upper critical field anisotropy $\gamma_H$ vs temperature
$T$, for unsubstituted, aluminum substituted, and carbon
substituted MgB$_2$. The results from 6.3\% C substituted (single
crystalline) MgB$_2$ are from Ref.\ \onlinecite{Puzniak04}.}
\label{Fig4}
\end{figure}

The development of $H_{c2}$ with partial carbon substitution is
also shown in Fig.\ \ref{Fig3} (open symbols). Our results, which
agree qualitatively with the limited results on single crystals
available,\cite{Lee03,Masui04,Pissas04,Puzniak04,Kazakov04} show a
drastic increase of $H_{c2}$ both parallel and perpendicular to
the layers, in contrast to the Al substitution case. The increase
of $H_{c2}^{\|c}$ is monotonous in the substitution range
measured, whereas $H_{c2}^{\|ab}$ peaks somewhat below $400\,{\rm
kOe}$ for C substitution levels around 5\%. Figure \ref{Fig4}
displays the temperature dependent $H_{c2}$ anisotropy for C
substituted, Al substituted, and unsubstituted MgB$_2$. Where
$H_{c2}^{\|ab} > 140\,{\rm kOe}$, $H_{c2}^{\|ab}$ results from
Refs.\ \onlinecite{Holanova04} and \onlinecite{Wilke04} have been
utilized to calculate $\gamma_H (T)$. With the exception of
Mg(B$_{0.9}$C$_{0.1}$)$_2$ and Mg$_{0.8}$Al$_{0.2}$B$_2$, where
the $T$ dependence of $\gamma_H$ is not obvious, a substantial
decrease of $\gamma_H$ with increasing $T$ is seen for all
samples. In accordance with the explanation of the $\gamma_H (T)$
dependence in unsubstituted MgB$_2$ this indicates that two band
effects are still relevant for both Al and C substituted MgB$_2$
for moderate substitution levels. The preservation of two distinct
superconducting gaps has indeed been observed directly on both Al
and C substituted
MgB$_2$.\cite{Ribeiro03,Putti03,Samuely03,Schmidt03,Putti04,Holanova04b,Gonnelli04b}
It is expected for moderate substitution levels from band
structure calculations, but also implies that interband scattering
cannot be substantially increased by partial substitutions of
either of these elements. The anisotropy monotonically decreases
with increasing substitution level, down to about $2$ for the
samples with the highest levels of substitution studied. This
decrease is rather similar for substitutions by Al and C. This
indicates that as far as the $H_{c2}$ {\em anisotropy} is
concerned, the main effect of carbon substitution is (as in the
case of Al substitution) a decreased anisotropy of $v_{F,\sigma}$
originating from the shifted Fermi level. The $T_c$ depression
(Fig.\ \ref{Fig1}) is also similar for Al and C substitution,
suggesting that this too may originate mainly from the shift of
$E_F$ and similar changes of phonon modes.

However, the drastically different $H_{c2}$ {\em magnitude} in C
substituted samples cannot be explained within this picture. The
difference to the Al substitution case is far too large to be
accounted for by different behavior of the lattice constants or
phonon modes, particularly given the above similarities in $T_c$
and $\gamma_H$ vs electron doping level. Rather, the very strong
increase of $H_{c2}$ points to the relevance of scattering in the
case of carbon substitution. Due to the two band nature of
superconductivity, three different scattering rates have to be
taken into account: interband scattering and intraband scattering
in the $\sigma$ and $\pi$ bands.\cite{Mazin02} We note that the
observation of constant $\Delta_{\sigma}/\Delta_{\pi}$ ratios by
spectroscopic means\cite{Holanova04b} indicate that the interband
scattering rate is hardly affected by moderate levels of C
substitution and may be neglected as in unsubstituted MgB$_2$.
However, a recent point contact study on MgB$_2$ crystals
containing high C substitution levels suggest that the interband
scattering rate may be important, particularly for high
substitution levels.\cite{Gonnelli04} Calculations within
``intraband dirty limit''\cite{Gurevich03,Golubov03} can indeed
explain very drastic increases of $H_{c2}$, much larger than in
single band superconductors.

These dirty limit calculations also yield a temperature dependent
$H_{c2}$ anisotropy, like the clean limit calculations do.
However, here the $T$ dependence of $\gamma_H$ also depends on the
ratio of the scattering in the anisotropic $\sigma$ and in the
nearly isotropic $\pi$ bands. If the intraband scattering is much
larger in the $\sigma$ bands than in the $\pi$ bands, a decreasing
$\gamma_H (T)$ dependence is expected, whereas the opposite case
results in an increasing $\gamma_H (T)$ dependence. Starting from
the unsubstituted MgB$_2$ with a decreasing $\gamma_H (T)$
dependence, a low level partial substitution mainly increasing the
$\sigma$($\pi$) bands scattering, should lead to a more(less)
pronounced $\gamma_H (T)$ dependence. As we can see from Fig.\
\ref{Fig4}, the $\gamma_H (T)$ dependence becomes less pronounced
upon increasing the substitution level. Comparing C and Al
substitution, the decrease of the $T$ dependence seems rather
similar, indicating that it is mostly due to the intrinsic changes
discussed above, rather than disorder. For the same electron
doping levels, the $\gamma_H (T)$ variation is somewhat less
strong for the carbon substitution case. This indicates that upon
C doping the scattering is increased more in the $\pi$ bands than
in the $\sigma$ bands. A similar conclusion was reached for a
6.3\% C substituted single crystal\cite{Puzniak04} and for thin
films containing carbon.\cite{Gurevich04,Braccini04} For a more
quantitative analysis, a theory treating clean-limit (electron
doping) and dirty-limit (scattering) effects on an equal footing
would be highly desirable.

The effect of higher $\pi$ band scattering also manifests itself
in the form of the $H_{c2}$ curves, most visibly for $H \| c$. In
extreme cases this leads to a positive curvature of
$H_{c2}^{\|c}(T)$ at low $T$.\cite{Gurevich03,Gurevich04} In
contrast to ``dirty films'' results\cite{Gurevich04,Braccini04} we
did not observe such a positive curvature, but compared to
unsubstituted MgB$_2$, the negative curvature of $H_{c2}^{\|c}(T)$
was significantly decreased for the C substituted samples. For
10\% C substituted MgB$_2$, $H_{c2}^{\|c}(T)$ was found to be
almost linear at low temperatures. The tendency of decreased
negative curvature of $H_{c2}^{\|c}$ with increasing C
substitution is also seen in single crystal
measurements\cite{Masui04,Puzniak04,Kazakov04} and supports the
conclusion of mainly additional scattering in the $\pi$ bands
causing the $H_{c2}$ enhancement. In contrast, in the Al
substituted samples, the $H_{c2}(T)$ curvature is not
significantly affected (cf.\ Fig.\ \ref{Fig2}), again indicating
less $\pi$ band scattering.

To account for an upper critical field that is much larger in C
substituted MgB$_2$, C substitution has to increase scattering in
the $\pi$ bands more relative to the $\sigma$ bands, and much more
than Al substitution does. That a substitution within the boron
layers would increase scattering more than a Mg site substitution
is hardly a surprise. It is, however, surprising that the increase
in the scattering is predominantly in the intraband scattering in
the isotropic $\pi$ bands. In partly C substituted thin films, the
effects of increased $\pi$ band scattering are much larger (as
visible both in the $H_{c2}(T)$ curve forms and in the $T$
dependence of the anisotropy). This scattering has been attributed
to a buckling of the $ab$ planes, tentatively due to nanophase
precipitates.\cite{Gurevich04,Braccini04} However, such
precipitates are unlikely to be present in our polycrystalline
samples or in single crystals, and we therefore conclude that an
increase of the $\pi$ bands scattering is an intrinsic property of
C substitution. This is in contrast to the aluminum substitution
case, where considerable variations of the significance of $\pi$
bands scattering exist for different
samples.\cite{note_Carrington} A recent first principles
electronic structure study\cite{Kasinathan04} on C substituted
MgB$_2$, taking into account disorder effects, found a larger
reduction of the mean free path in the $\sigma$ bands, which is in
contrast to our analysis. However, there are a variety of effects
that are more involved to include in a calculation, e.g., carbon
induced local distortions in the structure, as suggested by a
single crystal x-ray diffraction study.\cite{Kazakov04} Since the
large increase of $H_{c2}$ in C substituted MgB$_2$ and related
scattering rates are important for potential applications,
additional theoretical studies are clearly desirable, as would be
a clear experimental demonstration of a procedure boosting
scattering mainly in the $\sigma$ bands.

%\section{Conclusions}
\label{conc} In conclusion, whereas the development of $H_{c2}$
with partial aluminum substitution can be understood within a
simple rigid band picture, disorder effects are responsible for
the large enhancement of the $H_{c2}$ magnitude of carbon
substituted MgB$_2$. In contrast, the development of the $H_{c2}$
anisotropy and $T_c$ are remarkably similar for the two
substitutions. Scattering within the $\pi$ bands is increased much
more by carbon than by aluminum substitution, and more than
scattering in the $\sigma$ bands. The origin of this is yet to be
resolved.

%\begin{acknowledgments}
\label{ack} We thank V.~P. Antropov, V.~G. Kogan, and S.~A. Law
for useful discussions. Ames Laboratory is operated for the U.S.
Department of Energy by Iowa State University under Contract
W-7405-Eng-82. This work was supported by the Director for Energy
Research, Office of Basic Energy Sciences. MA gratefully
acknowledges financial support by the Swiss National Science
Foundation.
%\end{acknowledgments}

%\bibliographystyle{prstylong}
%\bibliography{MgB2,da_hist,da_pin,da_th_em,da_th_mi,da_mate,own,Y124}

\newcommand{\noopsort}[1]{} \newcommand{\printfirst}[2]{#1}
  \newcommand{\singleletter}[1]{#1} \newcommand{\switchargs}[2]{#2#1}

\end{document}